\documentclass[conference]{IEEEtran}
\IEEEoverridecommandlockouts
\usepackage[T1]{fontenc}
\usepackage{acronym}
\usepackage{cite}
\usepackage{amsmath,amssymb,amsfonts}
\usepackage{algorithmic}
\usepackage{graphicx}
\usepackage{textcomp}
\usepackage{xcolor}
\usepackage{lipsum}
\usepackage{cleveref}
\usepackage{siunitx}
\usepackage{xfrac}
\DeclareSIUnit \mAh {mAh}

\usepackage[hyphens]{url}
\def\BibTeX{{\rm B\kern-.05em{\sc i\kern-.025em b}\kern-.08em
    T\kern-.1667em\lower.7ex\hbox{E}\kern-.125emX}}

\usepackage{enumitem}
\usepackage{caption}
\usepackage{subcaption}
\usepackage{pgfplots}
\usepackage{pgfplotstable}
\pgfplotsset{compat = newest}

\usepackage{booktabs}

\graphicspath{{./img/}}

\acrodef{IFA}{\emph{Inverted F Antenna}}
\acrodef{BLE}{\emph{Bluetooth Low Energy}}
\acrodef{IoT}{\emph{Internet of Things}}
\acrodef{IETF}{\emph{Internet Engineering Task Force}}
\acrodef{SIG}{\emph{Special Interest Group}}
\acrodef{RFID}{\emph{Radio Frequency Identification}}
    
\definecolor{mblue}{RGB}{65,105,225}
\definecolor{mgreen}{RGB}{46,139,87}
\definecolor{mred}{RGB}{220,20,60}    
\definecolor{myellow}{HTML}{d9ac6c}

\begin{document}


\title{Optimizing Packet Reception Rates for Low Duty-Cycle BLE Relay Nodes
\thanks{This work is financed by the ERDF – European Regional Development Fund through the Operational Programme for Competitiveness and Internationalisation - COMPETE 2020 Programme within project WHERE.IS (POCI-01-0247-FEDER-024191).}
}

\author{\IEEEauthorblockN{Nuno Paulino, Luís M. Pessoa}
\IEEEauthorblockA{INESC TEC and Faculty of Engineering,\\University of Porto, Porto, Portugal\\
\{nuno.m.paulino, luis.m.pessoa\}@inesctec.pt}
\and
\IEEEauthorblockN{André Branquinho, Rafael Tavares, Igor Ferreira}
\IEEEauthorblockA{\textit{Wavecom}, Aveiro, Portugal \\
\{abranquinho, rtavares, iferreira\}@wavecom.pt}
}

\maketitle

\begin{abstract}
In order to achieve the full potential of the Internet-of-Things, connectivity between devices should be ubiquitous and efficient. Wireless mesh networks are a critical component to achieve this ubiquitous connectivity for a wide range of services, and are composed of terminal devices (i.e., nodes), such as sensors of various types, and wall powered gateway devices, which provide further internet connectivity (e..g, via WiFi).

When considering large indoor areas, such as hospitals or industrial scenarios, the mesh must cover a large area, which introduces concerns regarding range and the number of gateways needed and respective wall cabling infrastructure. Solutions for mesh networks implemented over different wireless protocols exist, like the recent \ac{BLE} 5.1. Besides range concerns, choosing which nodes forward data through the mesh has a large impact on performance and power consumption. 

We address the area coverage issue via a battery powered \ac{BLE} relay device of our own design, which acts as a range extender by forwarding packets from end nodes to gateways. We present the relay's design and experimentally determine the packet forwarding efficiency for several scenarios and configurations. In the best case, up to \SI{35}{\percent} of the packets transmitted by 11 nodes can be forwarded to a gateway by a single relay under continuous operation. A battery lifetime of 1 year can be achieved with a relay duty cycle of \SI{20}{\percent}.
\end{abstract}


\section{Introduction}
\label{sec:intro}

Wireless mesh networks can be the platform for many applications. A common use case are sensor networks \cite{darroudi2017bluetooth}, but others include domotics \cite{5473869}, automated inventory tracking or localization \cite{8692423}. Specific scenarios include healthcare \cite{Kamel2012,8531110}, security \cite{8270589,6583833} and warehouses and industrial facilities \cite{s20236766}.

Depending on the application, mesh networks can be built with WiFi devices, for example, but WiFi end-points or routers typically require wall power.  On the other hand, \ac{BLE} devices benefit from a comparatively lower cost, power efficiency, and smaller device sizes. The lower power consumption alone negates the need for installation of a wired infrastructure in favor of battery power, reducing costs further. A \ac{BLE} mesh network with battery powered end and relay nodes can be deployed in legacy locations without such an infrastructure. However, the range and data-rate for \ac{BLE} devices is lesser than that of protocols such as WiFi \cite{8371230}, meaning that denser networks may be required, which introduces the need for efficient are coverage and packet relaying.

We address the specific case where end nodes periodically transmit data on advertising channels only. The range of the network is restricted by the range of the edge nodes to the gateways, and on the available wall power for the gateways. 

To address this, intermediate nodes acting as relays are designed to extend the advertising range of the nodes. These nodes should also be battery powered, since otherwise they could be easily replaced with gateways. However, continuous operation by the relays to listen for sporadic transmissions from the nodes would result in an unsuitably short battery life.

By configuring their listening period with a low duty-cycle (i.e., configuring the network nodes to a low-power operating mode for a given listening period), the battery life can be considerably extended. Consequently, since \ac{BLE} packet transmission is sporadic, especially as the network nodes operate asynchronously, idling some (or all) the relay nodes inevitably leads to packet losses. However, some applications may not consider that all data is high priority, and some degree of data loss and/or end-to-end delay may be acceptable.

To characterize the efficiency of a system reliant on battery-powered relay nodes, we present an in-house design for such a \ac{BLE} relay, and characterize the system's packet loss in different conditions. Specifically, we vary the number of client nodes, the listening time spent on each \ac{BLE} channel, and apply two different forwarding policies, one of which has additional configuration parameters. Additionally, we subject the system to noise from other Bluetooth devices external to the network.  

We validate the operation of our \ac{BLE} relay design by manufacture and assembly, employing some of the units as beacons (so we may configure transmission periods), while another unit performs the relay function under several software configurations which implement our operating policies.

This paper is organized as follows: \Cref{sec:rwork} reviews related work, \Cref{sec:network} describes the network topology we addressed, \Cref{sec:node} presents the design characteristics of the \ac{BLE} relay node, and the configurable operating parameters, like the duty cycle and forwarding policy. \Cref{sec:experiment} presents experimental evaluation of packet reception rates for different scenarios. \Cref{sec:conclusion} concludes the paper. 

\section{Related Work}
\label{sec:rwork}

A comprehensive survey on the research efforts in \ac{BLE} mesh topologies is presented by Darroudi \emph{et al.} \cite{darroudi17}. The survey categorizes and compares nearly 30 approaches to \ac{BLE} network designs, including standardization solutions proposed by the Bluetooth \ac{SIG} and the \ac{IETF}, academic solutions, and proprietary solutions. 

A major distinction between mesh approaches is whether data is transmitted by flooding (e.g., using the \ac{BLE} advertising channels), or by through end-to-end connections through specific nodes. A comparison is presented in \cite{8292705}, where the authors compare the Trickle flooding protocol \cite{RFC6206} with the FruityMesh connection based protocol \cite{fruity}. Both are evaluated regarding their multi-hop efficiency, for a network of nine intermediate nodes placed between two source and sink nodes. The packet delivery ratio and the end-to-end delay are measured. Both approaches are comparable in this scenario, with a packet delivery rate of close to \SI{40}{\percent} when 10 packets are generated per second by the source node. FruityMesh suffers an end-to-end delay which is approximately 9x higher compared to Trickle, but in turn requires 3x less power.

Kim \emph{et al.} \cite{7300867} present \emph{BLEMesh}. A packet forwarding protocol is proposed to transmit batches of packets. Less transmissions are required in total to transport data end-to-end, through intermediate nodes, relative to naive flooding or routing based approaches. The packets include priority tables used by intermediate nodes to determine if a received packet should be re-transmitted, based on whether or not that packet was already forwarded by a node of higher priority. A downside is that the payload capability of the BLE packet diminishes as the number of nodes and batch size increases. A simulated evaluation for a mesh with 5 nodes, and assuming only one advertising channel, achieves a reduction of \SI{54.5}{\percent} in the required number of transmissions, relative to flood routing. 

Brandão \emph{et al.} \cite{9091162} propose the Drypp protocol, based on the Trickle flooding protocol \cite{RFC6206}. Trickle is a mesh network protocol for BLE where each node captures and attempts to re-transmit data at a later time, unless it meanwhile listens to redundant transmissions sent by other nodes. Drypp introduces a load balancing method which relies on dynamic adaptation of the protocol parameters based on each node's battery level. For three test nodes implementing the Drypp protocol, an \SI{11}{\percent} increase in network lifetime was achieved relative to Trickle, in exchange for a \SI{7.5}{\percent} decrease in throughput. 

A \ac{BLE} mesh network relying on a routing protocol is evaluated in \cite{asmir16}. The proposed mesh network is designed for environmental monitoring and disaster scenarios, and both the edge (sensor nodes) and the Wi-Fi capable gateway nodes are battery powered. Information if propagated based on Trickle routing \cite{RFC6206}. To extend battery life, the sensor nodes are periodically shut off, and modifications to the trickle algorithm are introduced to prevent packet loss due to these power-down periods. Given the periods for listening and transmission time, the authors estimated a lifetime of 589 days for a sensor node, and 511 days for a gateway, when equipped with \SI{6000}{{\milli\ampere\hour}} and \SI{8000}{{\milli\ampere\hour}} lithium polymer batteries, respectively. 

\begin{table*}[t]
\centering
\newcolumntype{R}[1]{>{\raggedleft\let\newline\\\arraybackslash\hspace{0pt}}m{#1}}
\caption{Summary of Experimental Results of Comparable Approaches}
\begin{tabular}{c|llR{0.9cm}R{1.1cm}R{1.1cm}R{1.1cm}R{1.4cm}R{2.3cm}R{1.5cm}}
\toprule
 Work & Protocol/Strategy & Type\textsuperscript{\textasteriskcentered} & Total \#Nodes & \#Sources / \#Sinks & Node Dist. (m) & TX Rate (p/second) & End-to-End Delay (\si{\milli\second}) & Packet Delivery Ratio (\%) & Node Power (\si{\milli\watt}) \\ \midrule
\cite{8292705} & FruityMesh & phys. & 9 & 1 / 1 & 1.5 & 1, 5, 10 & $\sim3.8E3$ & 100\%, $\sim$90\%, 40\% & 9.4\\
\cite{8292705} & Trickle    & phys. & 9 & 1 / 1 & 1.5 & 1, 5, 10 & $0.5E3$ & 100\%, 80\%, $\sim$35\% & 28.5 \\
\cite{9091162} & Drypp      & phys. & 5 & 1 / 1 & 1.5 & $\sim24$\textsuperscript{\textsection} & - & 91\%\textsuperscript{\textbar}  & 21.11\textsuperscript{\textbardbl}\\
\cite{7300867} & BLEMesh (batching) & sim. & 5 & 1 / 1 & - & - & - & -  & - \\
\cite{asmir16} & Trickle (modified) & phys. & 6\textsuperscript{\textdagger}  & 6 / 1 & $\sim$5 & 1/60 & 5--10 & - & 1.85\\
\cite{9140584} & MOC-CDS & sim. & 77--572 & 1 / 1 & 7 & 20--33\textsuperscript{\textsection} & 50--90 & $\sim$80\%--95\% & \textit{N/A}\textsuperscript{\textdaggerdbl} \\
\cite{9140584} & Genetic algorithm & sim. & 77--572 & 1 / 1 & 7 & 20--33\textsuperscript{\textsection} & 60--130 & $\sim$20\% & \textit{N/A}\textsuperscript{\textdaggerdbl} \\ 
\cite{bkeben2020efficient} & Minimum Relay Tree &sim. & 50\textsuperscript{\textdagger} & 50 / 1 & 20 & $\sfrac{5}{60}$, $\sfrac{2}{6}$ & 97, 99 & 63\%, 56\% & 1.5, 5.5 \\
\cite{bkeben2020efficient}  & Full Flooding & sim. & 50\textsuperscript{\textdagger} & 50 / 1 & 20 & $\sfrac{5}{60}$, $\sfrac{2}{6}$ & 102, 128 & 90\%, 82\% & 6.2, 22.6   \\
Ours & Immediate Fwd. & phys. & 4, 13 & 2, 11 / 1 & $\sim$1--5 & 1 & - & 58\%, 16\% & 24.74 \\
Ours & Batching and Fwd. & phys. & 13 & 11 / 1 & $\sim$1--5 & 1 & - & 35\%--9\% & 24.75--6.19 \\
\bottomrule
\multicolumn{10}{l}{} \\[-0.2em]
\multicolumn{10}{l}{\footnotesize{\textsuperscript{\textasteriskcentered}\emph{Simulation or Physical}, \textsuperscript{\textdagger}\emph{+1 Gateway}, \textsuperscript{\textdaggerdbl}\emph{Given only as relative decrease}, 
\textsuperscript{\textsection}\emph{To the best of our understanding}}} \\
\multicolumn{10}{l}{\footnotesize{\textsuperscript{\textbar}\emph{Relative to Trickle}, \textsuperscript{\textbardbl}\emph{Derived from reported values}}} \\
\end{tabular}
\label{tab:compare}
\end{table*}

The work in \cite{9140584} specifically addresses optimization of the use of Bluetooth relays in mesh networks. Connection-less mesh networks propagate data by controlled flooding between nodes, until the destination node of a particular data packet is reached. However, this leaves the network vulnerable to excessive flooding as a function of the number of nodes used as relays and/or selected to be relays. The authors employ state-of-the-art relay selection algorithms to a \ac{BLE} mesh network, and evaluate the effect of six different relay selection algorithms to a Connected Dominating Set (CDS) representation of the mesh. Using an in-house simulator, different relays densities were tested with two end nodes exchanging 1000 messages one-way. The lowest packet loss can be achieved by computing the routing with the fewest hops, but the lowest power consumption is possible for a genetic algorithm which find the minimum CDS of the network, at the cost of suffering the highest packet loss (as high as \SI{80}{\percent}). 

In \cite{bkeben2020efficient}, a method for relay node management is proposed based on a tree representation for the mesh network, together with an integer linear programming formulation which minimizes the number of relay nodes required to ensure connectivity between all nodes. The algorithm requires that the number of nodes and network topology be know to determine the relay routing. Using an in-house simulator, the authors evaluate the routing efficiency and energy consumption of a system composed of up to 100 nodes in an indoor configuration where line-of-sight is not possible for all pairs of nodes. A power consumption reduction of up to 12x is claimed over the conventional case where any relay node can be used as a relay during forwarding (i.e., flooding).

In general, the choice of protocol and network topology is application dependant. \Cref{tab:compare} summarizes the results (or a subset of results) from the experimental evaluations shown in this section. The values reported are our best effort at a comparison of the presented approaches, as well as our own. Depending on the respective experiments, some columns show either scalar, ranges of values, or lists (correspondence between list values is kept column to column). Node power reports the power consumption of each node of the tested mesh, taking into account the entire operating time, including any sleep periods of the nodes (i.e., the average power consumption throughout the experiment lifetime.

The experiments we conducted can be categorized as controlled flooding mechanism, but where we rely on details specific to a class of applications to determine forwarding behaviour. We consider end nodes with a constant packet rate, and envision a tree topology for the network where a relay is responsible for the end nodes within its range, and where relays are out of range amongst themselves. Additionally, we are not concerned with end-to-end delay, as data is non-critical given equal importance. We also conduct experiments while introducing real-world noise due to other wireless devices external to the network, which we have not observed in other works we have identified. 

\section{Network Topology}
\label{sec:network}

\begin{figure}
\centering
\includegraphics[width=0.58\linewidth]{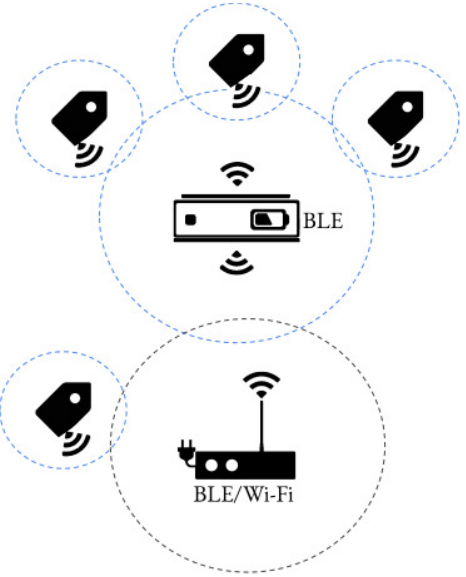}
\caption{BLE mesh topology, with battery-powered end nodes and intermediate relay nodes, and wall-powered BLE/Wi-Fi gateways interfacing with an upstream server system}
\label{fig:network}
\end{figure}

The use-case network topology for the evaluation of our relay, and respective forwarding policies, is shown in \Cref{fig:network}. We target use cases where the end nodes are battery powered, and periodically transmit information about the environment (e.g., sensor data). The gateways are \ac{BLE}/Wi-Fi devices which synchronize the status of the network with the centralized system. The network was designed and tested according to the features/constraints of the Bluetooth 4.1 specification \cite{ble41}.

One of the characteristics of \ac{BLE} is the transmission range (approximately \SI{20}{\meter}). This means that either all nodes placed throughout the site have to be within this range of a wall-powered gateway in order for data to be retrieved by those nodes, or that data is forwarded through nodes. The former is a potentially expensive solution, and the later is the object of study on mesh network routing protocols. 
However, if the end-nodes are simple sensors and cannot move data to and from each other (or if they are physically placed in such a way that a sequence of hops from end node to gateway cannot be established), more sophisticated battery-powered intermediate nodes are required which do not gather data themselves, but serve as range extenders to the gateways. 

This paper focuses presents a design of the relay node, which functions as a packet receiver, gatherer, and re-transmitter. This makes it possible to extend the network range in situations where the indoor configuration or cost do not allow for a more ubiquitous distribution of wall-powered gateways. It also provides a cheaper solution relative to fully-fledged gateways, since it may replace them where Wi-Fi capabilities are not needed. Additionally, since the relays are battery powered, they are easy to relocate according to changes in the application requirements, or simply to tune the quality of the sensed data. The relays are compatible with any off-the-shelf end node which is BLE/Bluetooth 4.1 compliant. 

\section{\ac{BLE} Relay Node}
\label{sec:node}

The purpose of the \ac{BLE} relay device is to serve as a packet forwarder. It discards (i.e, does not forward) packets originating from devices which are not part of the its own network. Currently this is done by MAC address filtering. The only payload sent is the identification of each node. 

We implemented two relay designs, both based on a single Nordic Semiconductor \emph{nRF52832} micro-controller \cite{nordic1}, which performs the packet reception and re-transmission, and idles the relay by going into a low-power mode. The configuration parameters listed earlier, such as listening intervals and periodicity, are controlled by the firmware residing on the non-volatile program memory of the \emph{nRF52832} chip. All relay implementations are composed by one single-layer, dual-sided, FR-4 PCB with a \SI{1}{\milli\meter} thickness. 

\begin{figure}
     \centering
     \begin{subfigure}[b]{1\linewidth}
        \centering
        \includegraphics[width=0.48\linewidth]{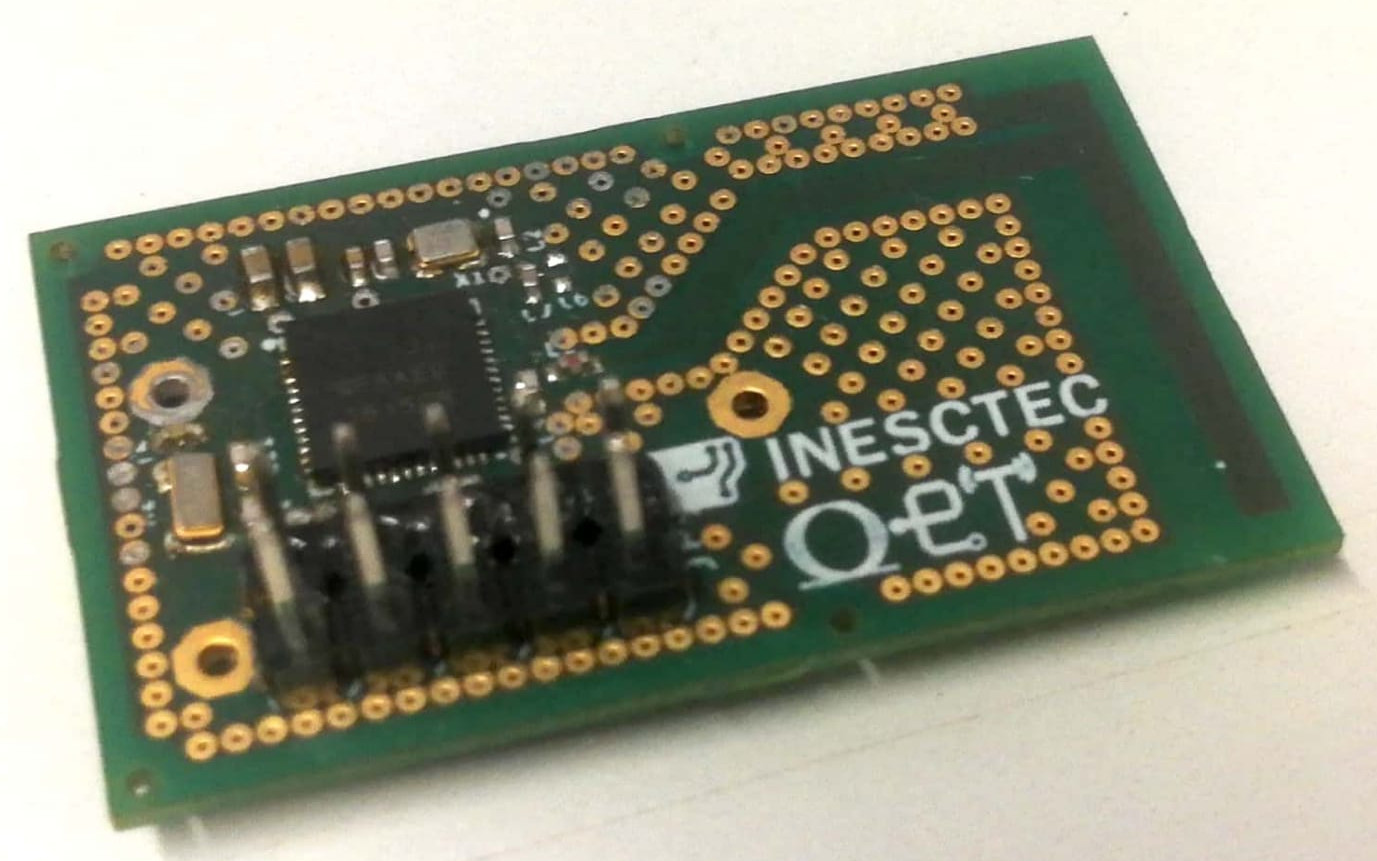}
        \caption{BLE relay prototype A, powered by a \SI{3.3}{\volt} button cell}
        \label{fig:protoA}
     \end{subfigure}
     \hfill
     
     \begin{subfigure}[b]{1\linewidth}
         \centering
        \includegraphics[width=.521\linewidth]{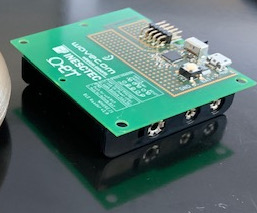}
        \includegraphics[width=.35\linewidth]{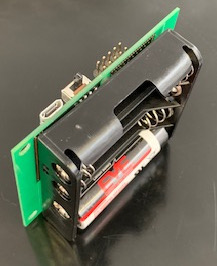}
        \caption{BLE relay prototype B, with multiple power sources; the experimental evaluations in this paper consider only this relay variant}
        \label{fig:protoB}
     \end{subfigure}
    \caption{Two variants of the relay prototype; functionally identical with different power sources}
    \label{fig:relays}     
\end{figure}

The first prototype relay is shown in \Cref{fig:protoA}. It contains the \emph{nRF52832} chip, a J-Link type programming header, and a single \SI{3.3}{\volt} CR2032 button cell battery. The relay is considerably small, with with a \SI{23x38x10}{\milli\meter} profile. The antenna for reception and transmission of Bluetooth packets is a co-planar \ac{IFA}, tuned for \SI{2.4}{\giga\hertz}. 

A second prototype designed for longer lifespan is shown in \Cref{fig:protoB}. A series of four \SI{3.3}{\volt} AA batteries powers the relay when deployed in a location where wall power is unavailable, which is the primary use-case of the device. Alternatively, a mini-USB connector accepts a \SI{5}{\volt} input. An \emph{LTC4419} chip \cite{ltc4419} is used as a power selector, which prioritizes the USB power input. A \emph{TPS62125} \cite{tps62125} regulates the chosen input to \SI{3.3}{\volt} for the \emph{nRF52832}. Finally, the J-Link programming header powers the device in the absence of other power sources. The antenna design is identical to that of prototype \emph{A} (albeit with a longer trace to the PCB edge, of \SI{2.1}{\centi\meter}), and the device is \SI{74x64x25}{\milli\meter}. 
\noindent\noindent\begin{figure}%
\centering%
\begin{tikzpicture}%
\begin{axis}[%
    xmin=0,
    xmax=60,
    ymin=0.5,
    ymax=2.5,
	xlabel = time (\si{\milli\second}),
	x label style={anchor=west},
	xticklabels=\empty,
	axis y line=none,
	xmajorgrids,
	xminorgrids,
	inner axis line style={latex-latex},
	axis lines=center,
	width=0.95\linewidth, scale only axis,
	height=0.30\linewidth,
	legend style={
	    draw=none,
	    at={(axis cs:0.0,0.0)},
        anchor=west,legend columns=3},
    legend style={
        /tikz/every even column/.append style={column sep=0.3cm}}
    ]
    
    \draw[line width=3mm, gray] (axis cs:30.5,2) -> (axis cs:50,2);
    
    \draw[line width=3mm, mblue] (axis cs:0,2) -> (axis cs:20,2);
    
    \draw[line width=3mm, myellow] (axis cs:20.5,2) -> (axis cs:30,2);

    \draw[line width=3mm, mgreen] (axis cs:0,1.5) -> (axis cs:20,1.5);    
    \draw[black, thick] (axis cs:0,1.3) -> (axis cs:0,1.7);
    \draw[black, thick] (axis cs:5,1.3) -> (axis cs:5,1.7);
    \draw[black, thick] (axis cs:10,1.3) -> (axis cs:10,1.7);
    \draw[black, thick] (axis cs:15,1.3) -> (axis cs:15,1.7);
    \draw[black, thick] (axis cs:20,1.3) -> (axis cs:20,1.7);
    

    
    \draw[line width=3mm, mred] (axis cs:0,1) -> (axis cs:2.5,1);
    \draw[line width=3mm, mred] (axis cs:5,1) -> (axis cs:7.5,1);
    \draw[line width=3mm, mred] (axis cs:10,1) -> (axis cs:12.5,1);

    \draw[line width=3mm, mred] (axis cs:15,1) -> (axis cs:17.5,1);
    
    \draw[line width=3mm, mblue!20!white] (axis cs:50.5,2) -> (axis cs:60,2);
    \draw[line width=3mm, mgreen!20!white] (axis cs:50.5,1.5) -> (axis cs:55.5,1.5);
    \draw[line width=3mm, mred!20!white] (axis cs:50.5,1) -> (axis cs:53,1);
    \draw[line width=3mm, mred!20!white] (axis cs:55.5,1) -> (axis cs:58,1);
    
    \addlegendimage{line width=0.4mm,color=gray}
    \addlegendentry{Sleep Time}
    
    \addlegendimage{line width=0.4mm,color=mblue}
    \addlegendentry{Scan Time}
    
    \addlegendimage{line width=0.4mm,color=myellow}
    \addlegendentry{Forwarding}
    
    \addlegendimage{line width=0.4mm,color=mgreen}
    \addlegendentry{Scan Interval}
    
    \addlegendimage{line width=0.4mm,color=mred}
    \addlegendentry{Scan Window}
    
    
    \end{axis}
\end{tikzpicture}
\caption{Temporal diagram of relay operation cycle. The length of the sleep time (radio off, and sleep mode enabled for the \emph{nRF52832}) determines the overall duty cycle. The \emph{forwarding} period applies for one of our forwarding policies (\Cref{sub:fwd}).}
\label{fig:cycle}
\end{figure}
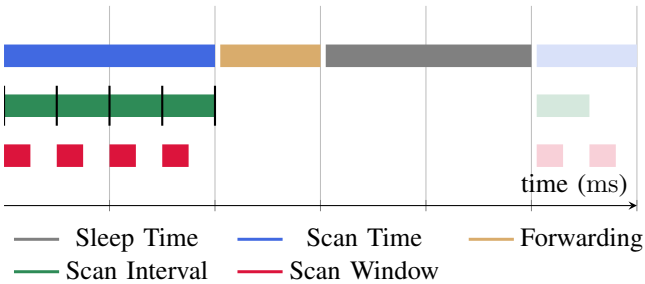
The relay's software can accept a number of configuration parameters which will be the focus of the experimental evaluation. \Cref{fig:cycle} shows the cyclical operation mode of the relay during scanning. The relay stays in a given channel during a \emph{scan interval}, and listens on that channel during the length of the \emph{scan window}. In our tests we vary the length of the scan interval and set the scan window to an equal value. We evaluate the effects of two forwarding policies and estimate lifetime of the devices as a function of the sleep time (for the best performing scan interval and policy). Only advertising channels are used, and paired connections are not established, which is typical for one-way sensor meshes. 

\section{Experimental Evaluation}
\label{sec:experiment}

\begin{figure}
\centering
\includegraphics[width=0.85\linewidth]{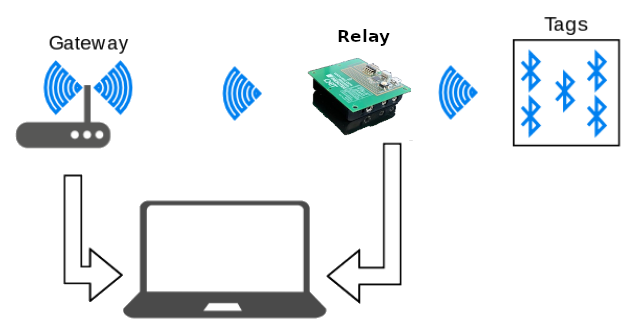}
\caption{Experimental setup for relay efficiency evaluation}
\label{fig:setup}
\end{figure}

We evaluate the relay's performance regarding packet reception and forwarding, for different scan interval lengths, policies, and sleep time. We employed the experimental setup show in \Cref{fig:setup}. In addition to the elements of the system shown, additional \ac{BLE} nodes were placed in the environment, to act as noise, thus subjecting the system to a realistic operating condition. For all our tests, the \emph{scan window} occupies the entire duration of the \emph{scan interval}, in order to evaluate only the effects of the listening time, forwarding policy and sleep time. Exploring the effects of the length of sleep time (i.e., device duty cycle), in conjunction with non-equal scan window and interval lengths, , on power savings and performance is out of the scope of this paper. 

Given this, we evaluated the following characteristics: 
\begin{itemize}
\item the rate of packets received by the relay while subject to noise, for different scan intervals (i.e., advertising channel switching periods);
\item the forwarding efficiency between the relay and a gateway using an immediate forwarding policy, first with two client nodes, and then with 11 client nodes; 
\item forwarding efficiency for 11 nodes, under a policy which buffers received packets and forwards replicas to the gateway, to reduce the overhead of switching between radio modes;
\item power consumption as a function of device duty cycle (i.e., sleep time).
\end{itemize}

In order to account for all transmitted and received packets, the relay and the terminal gateway communicate every packet received via serial connection. Each packet is annotated with the originating node. Since the transmission period of the nodes is known, we know the total transmitted packets for a given run time. We can then compute the packet losses in different conditions, between the nodes and the relay, and between the relay and the gateway. 

\subsection{Relay Reception Efficiency for 2 Nodes}

In this test, the relay's packet reception rate under noise was tested for two client nodes,  set to transmit advertising packets with period of \SI{1}{\second}. The test environment contained another 15 \ac{BLE} nodes, external to the network, advertising at different intervals and thus acting as noise. 

We varied the relay's \emph{scan interval} between \SI{50}{\milli\second} and \SI{1150}{\milli\second}. The \emph{scan window} occupies the entire period. 
What is measured in this case is the packet reception rate under noise, and due to the intrinsic loss of packets due to the randomness of the selected transmission and reception channels. The Bluetooth specification outlines a total of 40 channels, three of which (37, 38 and 39) are used for advertising packets. 

\noindent%
\noindent\begin{figure}%
    \centering%
        \begin{tikzpicture}%
        \begin{axis}[%
        yticklabel={\pgfmathparse{\tick*100}\pgfmathprintnumber{\pgfmathresult}\%},
        xtick=data,
        xticklabel style={rotate=45},
        height=0.60\linewidth,
        width=1\linewidth,
        grid = both,
        major grid style = {lightgray},
        minor grid style = {lightgray!25},
        xlabel=Relay Scan Interval (ms),
        ylabel=Relay Reception Rate (\%),
        legend pos = north west]

        \addplot[mblue, only marks] 
        table [x=scanwindow, y=efficiency, col sep=comma] {teste3.csv};

        \addplot[mred, thick] table[
          col sep=comma,
          x=scanwindow,
          y={create col/linear regression = {y=efficiency}}
        ] {teste3.csv};

        \addlegendentry{Nodes to Relay Rate}
        \end{axis}
        \end{tikzpicture}
\caption{Reception rate between the nodes and relay, and the relay and gateway, for three runs per relay scan interval (for 1200 packets sent by two nodes, transmitting at \SI{1}{\second} intervals), without forwarding, and 17 nodes acting as noise.}
    \label{fig:test3}
\end{figure}
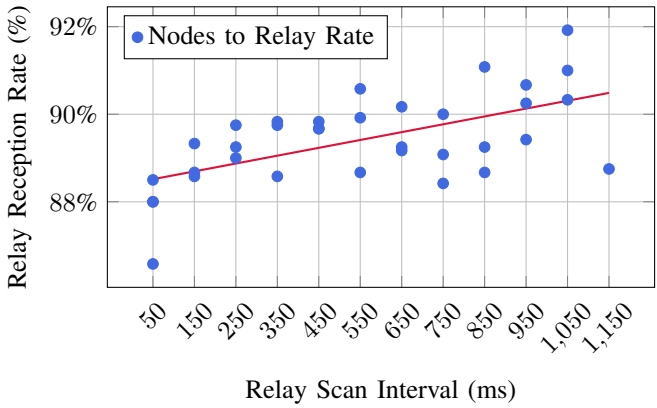

\Cref{fig:test3} shows the measured reception rates of the relay. Three runs were performed per configuration. Per run, each of the two nodes transmitted 600 packets. For a transmission rate of 1 packet per second, this totals an experimental time of \SI{90}{\minute} per configuration. For all experiments the average reception rate is \SI{88}{\percent} ($\sigma = 1.02\%$). 

The \emph{scan interval} does not affect the reception rate significantly. Even so, it is marginally more efficient for the relay to stay tuned into a single channel for as long as possible, i.e., longer scan intervals. This might contribute to a slightly reduced packet loss since less time is spent switching radio channels, which contributes to idle time. 
Also, since the Bluetooth protocol also dictates that an advertising event must be sent by a node on all three channels, the likelihood of the relay capturing a packet is higher by staying on a single channel for a period of time which is greater than the node's transmission period. 

Note that in this scenario the relay's radio never transmits, and we evaluated the best case reception rate in a noisy scenario. Since the radio is half-duplex, once the relay begins forwarding packets, its reception rate will consequently decrease, as we present next.

\subsection{Relay Forwarding Efficiency for 2 Nodes and 11 Nodes}

\Cref{fig:test4} shows the reception efficiencies between the nodes and the relay, and between the relay and gateway. \Cref{fig:test4a} shows the case with 2 client nodes, and 15 nodes acting as noise, and \Cref{fig:test4b} shows the case with 11 client nodes, and 6 nodes acting as noise.
\noindent%
\pgfplotsset{
mystylea/.style={
yticklabel={\pgfmathparse{\tick*100}\pgfmathprintnumber{\pgfmathresult}\%},
xtick=data,
xticklabel style={rotate=45},
height=0.60\linewidth,
width=1\linewidth,
grid = both,
ymax=0.8,
ymin=0,
major grid style = {lightgray},
minor grid style = {lightgray!25},
xlabel=Relay Scan Interval (ms),
ylabel=Reception Rate (\%),
legend pos = north east}
}
\noindent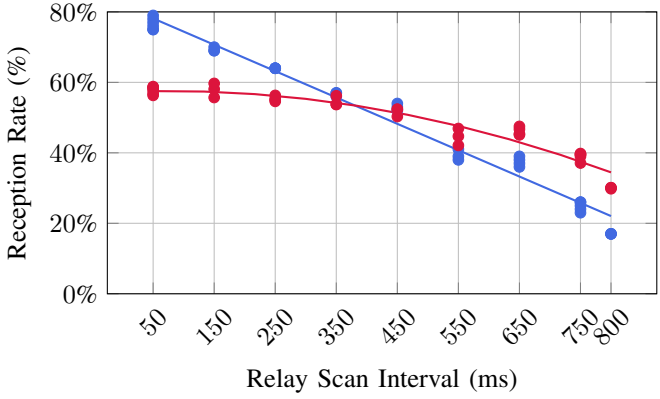
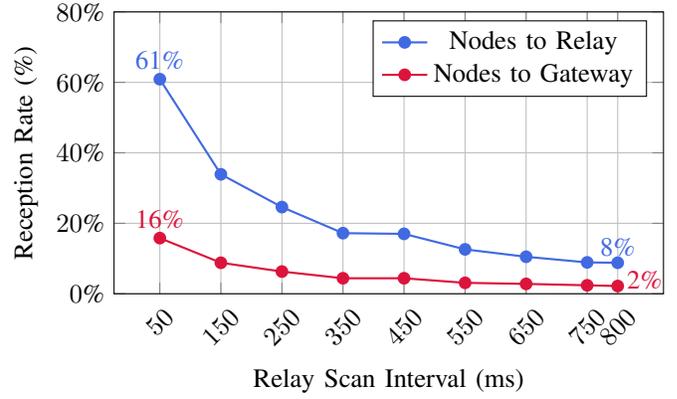
\begin{figure*}%
    \centering%
    \begin{subfigure}[t]{0.49\linewidth}
        \begin{tikzpicture}%
            \begin{axis}[mystylea]

                \addplot[only marks, mblue] table 
                [x=scanwindow, 
                y=efficiencyTagRelay, 
                col sep=comma] {teste4.csv};
                
                \addplot[only marks, mred] table 
                [x=scanwindow, 
                y=efficiencyTagGateway, 
                col sep=comma] {teste4.csv}; 
                
                \addplot[thick, mblue] table 
                [ col sep=comma, x=scanwindow,
                  y={create col/linear regression = {y=efficiencyTagRelay}}
                ] {teste4.csv};
                
                \addplot[mred, thick]  plot[samples=10,domain=50:800]  {0.573+6.57e-5*x-4.39e-07*x*x};
            
            \end{axis}
        \end{tikzpicture}
        \caption{relay and gateway reception rate for 2 nodes, for a total of 1200 packets sent by 2 nodes at \SI{1}{\second} intervals, with 17 nodes acting as noise}
        \label{fig:test4a}
    \end{subfigure}
    \hfill
    \begin{subfigure}[t]{0.49\linewidth}
        \begin{tikzpicture}%
        \begin{axis}[mystylea]
        
                \addplot[mblue, mark color=mblue,  mark=*, thick] table 
                [x=scanwindow, 
                y=efficiencyTagRelay, 
                col sep=comma] {teste8.csv};
                \node[anchor = south, mblue] at (50,0.61) {61\%};
                \node[anchor = south, mblue] at (800,0.08) {8\%};
                
                \addplot[mred, mark color=mred,  mark=*, thick] table 
                [x=scanwindow, 
                y=efficiencyTagGateway, 
                col sep=comma] {teste8.csv}; 
                \node[anchor = south, mred] at (50,0.16) {16\%};
                \node[anchor = west, mred] at (800,0.04) {2\%};
                
                \addlegendentry{Nodes to Relay}
                \addlegendentry{Nodes to Gateway}
        \end{axis}
        \end{tikzpicture}
        \caption{relay and gateway reception rate for 2 nodes, for a total of 39600 packets sent by 11 nodes at \SI{1}{\second} intervals, with 6 nodes acting as noise}
        \label{fig:test4b}
    \end{subfigure}      
\caption{Reception rate for the relay and gateway for two scenarios under an immediate forwarding policy}
\label{fig:test4}
\end{figure*}

In these experiments, the sleep time is zero, as we wish to evaluate the performance, for a long period of operation, only as a function of the network size, scan interval, and noise introduced by other devices. The relay has an immediate forwarding policy for every packet received. 

\Cref{fig:test4a} shows that the relay experiences a greater packet loss relative to the data in \Cref{fig:test3}, since it was configured to interrupt the scan interval and re-transmit immediately. This policy intended to reduce the travel time of the packets to the gateway. However this means that only one packet is relayed per scan interval, which explains the loss of packets from the nodes to the relay. Consequently, the number of packets forwarded to the gateway diminishes as the scan interval increases. 

For scan intervals greater than \SI{350}{\milli\second}, the number of packets received by the gateways actually exceeds those forwarded. This is due to two factors. Firstly, for forwarding the relay must be switched to advertising mode for a duration such that only one packet is sent. However, non-deterministic behaviour during channel switching and switching between reception and transmission sometimes produces duplicate packets. Secondly, the gateway may receive packets directly from the nodes, depending on transmission power. This leads to an apparent increase in system performance for lengthier scan intervals, despite the relay's losses.

\Cref{fig:test4b} shows the same metrics when 11 nodes are introduced into the system. For the same reason as before, the reception rate (for both the relay and gateway) decreases with the relay's scan interval. However, this case shows how the relay effectively acts as an intermediate buffer to hold packets. The shorter the scan intervals, the quicker the relay echoes packets, decreasing the likelihood that packets are missed while the gateway is occupied, either by being in a non-listening state, e.g., switching between channels, or by being busy processing beacons received either directly from the nodes or by the relay.

However, even in the best case, only approximately \SI{16}{\percent} of the total packets sent arrive at the gateway, which implies significant energy expenditure by the beacons without benefit. The next section improves this with a different forwarding policy.

\subsection{Relay Forwarding Efficiency for 11 Nodes \& Batching Policy}
\label{sub:fwd}

We programmed the relay with a forwarding policy based on a listening period, and a forwarding period. During the listening period, the relay accumulates the captured packets, e.g., 4 packets from node \#1, 10 from node \#2, and one from node \#3. During forwarding, the relay echoes up to $N$ repetitions of a packet per node, regardless of how many packets were received per node. For instance, for 10 packets received for node \#1, five echoes will be transmitted. This reduces the total traffic, and also normalizes the amount of packets sent upstream to the gateways, potentially boosting reception of packets sent by nodes under noiser conditions. 

\noindent%
\pgfplotsset{
mystyle/.style={
yticklabel={\pgfmathparse{\tick*100}\pgfmathprintnumber{\pgfmathresult}\%},
xtick=data, xticklabel style={rotate=45},
height=1\linewidth, width=1\linewidth,
ymin=0,
ymax=1,
xmin=0,
grid = both, major grid style = {lightgray}, minor grid style = {lightgray!25},
xlabel=Repeat Interval (\si{\milli\second}),
ylabel=Reception Rate (\%), legend pos = north east}
}
\noindent\begin{figure*}%
    \centering%
    \begin{subfigure}[t]{0.31\linewidth}
        \begin{tikzpicture}%
            \begin{axis}[mystyle,
            legend style={
    	    draw=none,
    	    at={(axis cs:0.0,1.25)},
            anchor=north west,legend columns=4},
            legend style={
                /tikz/every even column/.append style={column sep=0.5cm}}
            ]
            \addplot[mred, very thick] 
            table [x=repeat, y=wake5, col sep=comma] {teste9.csv};
            \addplot[mblue, very thick] 
            table [x=repeat, y=rt5, col sep=comma] {teste9.csv};
            \addplot[mgreen, very thick] 
            table [x=repeat, y=gr5, col sep=comma] {teste9.csv};
            \addplot[myellow, very thick] 
            table [x=repeat, y=gt5, col sep=comma] {teste9.csv};
            \node[anchor = south, myellow] at (10,0.35) {35\%};
            \node[anchor = south, myellow] at (100,0.23) {23\%};
            \addlegendentry{Listen Time / Total Time (\%)}
            \addlegendentry{Nodes to Relay}
            \addlegendentry{Relay to Gateway}
            \addlegendentry{Nodes to Gateway}
            \end{axis}
        \end{tikzpicture}
        \caption{reception rates for 5 packet repetitions from relay to gateway}
        \label{fig:test9a}
    \end{subfigure}
    \hfill
    \begin{subfigure}[t]{0.31\linewidth}
        \begin{tikzpicture}%
            \begin{axis}[mystyle]
            \addplot[mred, very thick] 
            table [x=repeat, y=wake2, col sep=comma] {teste9.csv};
            \addplot[mblue, very thick] 
            table [x=repeat, y=rt2, col sep=comma] {teste9.csv};
            \addplot[mgreen, very thick] 
            table [x=repeat, y=gr2, col sep=comma] {teste9.csv};
            \addplot[myellow, very thick] 
            table [x=repeat, y=gt2, col sep=comma] {teste9.csv};
            \node[anchor = south, myellow] at (10,0.15) {15\%};
            \node[anchor = south, myellow] at (100,0.13) {13\%};
            \end{axis}
        \end{tikzpicture}
        \caption{reception rates for 2 packet repetitions from relay to gateway}
        \label{fig:test9b}
    \end{subfigure}    
    \hfill
    \begin{subfigure}[t]{0.31\linewidth}
        \begin{tikzpicture}%
            \begin{axis}[mystyle]
            \addplot[mred, very thick] 
            table [x=repeat, y=wake1, col sep=comma] {teste9.csv};
            \addplot[mblue, very thick] 
            table [x=repeat, y=rt1, col sep=comma] {teste9.csv};
            \addplot[mgreen, very thick] 
            table [x=repeat, y=gr1, col sep=comma] {teste9.csv};
            \addplot[myellow, very thick] 
            table [x=repeat, y=gt1, col sep=comma] {teste9.csv};
            \node[anchor = south, myellow] at (25,0.07) {7\%};
            \node[anchor = south, myellow] at (100,0.07) {7\%};
            \end{axis}
        \end{tikzpicture}
        \caption{reception rates for 1 packet repetition from relay to gateway}
        \label{fig:test9c}
    \end{subfigure}       
\caption{Reception rates between the several levels of the network, for different forwarding repetitions (i.e., \Cref{fig:test9a,fig:test9b,fig:test9c}), and for different intervals between repetitions. The \emph{listening time} indicates the time the relay spends listening to the nodesand the remaining time is spent forwarding. The scan interval and window are \SI{50}{\milli\second} for all cases.}
\label{fig:test9}
\end{figure*}
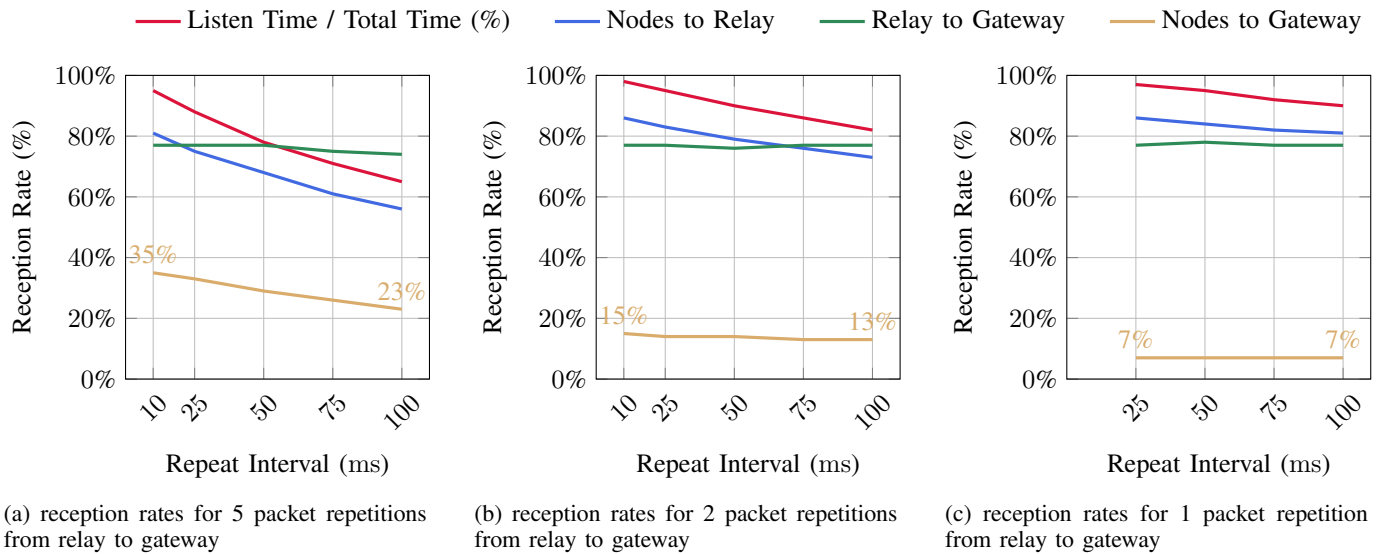

\Cref{fig:test9} shows the reception rates, this time including also the rate of successful transfer between the relay and gateway. The three scenarios employ a listening period of \SI{10}{\second}, and a different number of packet repetitions each, e.g., five packet repetitions for \Cref{fig:test9a}. For each case, the interval between repetitions is also varied. Once again, the sleep time is zero, and the \emph{scan interval} is \SI{50}{\milli\second} for all cases.

The \emph{listening time} is also shown, which represents the amount of time during each listen-and-forward cycle that the relay is listening. The relay first listens during the \emph{scan time} ($S_{Time}$) (switching between channels every \emph{scan interval}), and buffers the packets. Then it enters forwarding mode where each packet is re-sent a given number of times (\emph{$Nr_{Repeats}$}) at a set interval (\emph{$R_{Interval}$}). Given that there are 11 nodes, the ratio between listening and forwarding time can be estimated as:
\noindent\begin{equation}
L(\%) = \frac{S_{Time}}{S_{Time} + R_{Interval} \times Nr_{Repeats} \times N_{Nodes} }
\end{equation}

In the best case, with 5 repetitions at \SI{10}{\milli\second} interval, up to \SI{35}{\percent} of packets are now successfully forwarded to the gateway, which is 2.2x increase in performance relative to immediate forwarding. Although the relay captures less packets directly from the nodes, due to the lengthier forwarding period, the overall forwarding efficiency is higher. 

In a multi-relay scenario, a superior performance should be expected, although the best strategy regarding scan interval, repeat interval, and repeat count would have to be determined. However, a possible approach would be to have each relay in the system forward only a subset of all nodes, thus reducing its own load and preventing excessive in-system noise. 

\subsection{Estimated Power Consumption vs. Sleep Time}

This section explores the power consumption in continued operation as a function of sleep time, given that the forwarding rates during uptime are indicated by the previous experiments. To retrieve power consumption, we utilized a power profiler kit from Nordic Semiconductors \cite{nordic2}.

\noindent%
\pgfplotsset{
mystyle2/.style={
xticklabel={\pgfmathparse{\tick*100}\pgfmathprintnumber{\pgfmathresult}\%},
xtick=data,
xticklabel style={rotate=45},
grid=both,
major grid style = {lightgray},
height=0.60\linewidth,
width=0.90\linewidth}
}
\noindent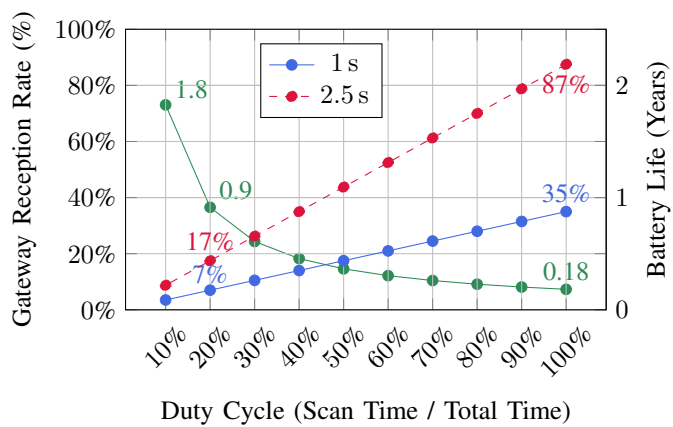
\begin{figure}%
    \centering%
        \begin{tikzpicture}%

        \begin{axis}[%
        mystyle2, 
        axis y line*=right, 
        axis x line=none,
        ymax=2.5, 
        ymin=0, 
        ylabel=Battery Life (Years)]
        
        \addplot[mgreen, mark=*]
        table [x=dutyCycle, y=battlifeY, col sep=comma] {power.csv};        
        \node[anchor = south west, mgreen] at (0.10,1.8) {1.8};
        \node[anchor = south west, mgreen] at (0.20,0.9) {0.9};
        \node[anchor = south, mgreen] at (1.00,0.18) {0.18};
        \end{axis}
        
        \begin{axis}[%
        mystyle2,
        yticklabel={\pgfmathparse{\tick*100}\pgfmathprintnumber{\pgfmathresult}\%},
        xlabel=Duty Cycle (Scan Time / Total Time),
        ylabel=Gateway Reception Rate (\%),
        ymax=1,
        ymin=0,
        legend style={
    	    at={(axis cs:0.45,0.95)},
            anchor=north}]
        \addplot[mblue, mark=*] 
        table [x=dutyCycle, y=1second, col sep=comma] {power.csv};
        \label{line1}
        \node[anchor = south, mblue] at (1.0,0.35) {35\%};
        \node[anchor = south, mblue] at (0.2,0.06) {7\%};
        \addplot[mred, mark=*, dashed] 
        table [x=dutyCycle, y=25second, col sep=comma] {power.csv};
        \label{line2}
        \node[anchor = north, mred] at (1.0,0.87) {87\%};
        \node[anchor = south, mred] at (0.2,0.18) {17\%};
        \addlegendentry{\SI{1}{\second}}
        \addlegendentry{\SI{2.5}{\second}}
        \end{axis}        
        
        \end{tikzpicture}   
\caption{Battery life and gateway reception rate as a function of duty cycle. Reception rate is shown for a system with 11 client nodes with a transmission period of \SI{1}{\second} and \SI{2.5}{\second}. The relay is configured for a \SI{50}{\milli\second} scan interval/window, and with a batching policy of 5 repetitions and \SI{10}{\milli\second} repeat time.}
\label{fig:power}
\end{figure}

We first use the power profiler to measure the current draw during radio operation (i.e., during \emph{scan window} periods). Regardless of configuration values, the relay draws \SI{7.5}{\milli\ampere}. 

We then evaluate the power consumption for different duty cycles defined by the \emph{scan} and \emph{sleep} times. The \emph{scan interval} and \emph{window} remain equal at \SI{50}{\milli\second}, and adopt a batching policy with 5 repetitions and a \SI{10}{\milli\second} repeat time. The efficiency for this case was \SI{35}{\percent}. 
The average current draw and efficiency as a function of the duty cycle can be calculated by the product of the cycle and these baseline values of \SI{7.5}{\milli\ampere} and \SI{35}{\percent}, respectively. The battery life is computed based on the relay's four AA batteries totaling \SI{12000}{\mAh}. 

\Cref{fig:power} shows the resulting efficiencies and battery life. The efficiency is shown based on the experimental runs with 11 nodes transmitting at a \SI{1}{\second} interval. In this case, a duty cycle of \SI{100}{\percent} leads to the \SI{35}{\percent} efficiency, but a battery life of only approximately 2.16 months. To attain a battery life of a year, a duty cycle of \SI{20}{\percent} is required, with an estimated efficiency of \SI{7}{\percent}. Note that the effective efficiency of forwarding remains \SI{35}{\percent}, since a duty cycle of \SI{20}{\percent} implies that, in the best case, \SI{20}{\percent} of all packets would be forwarded. 

Additionally, note that for all experiments, the efficiency is dependant on the total amount of packets sent by the nodes. These experimental runs impose a \SI{1}{\second} period per node. For some applications like sensor networks for temperature or light intensity readings with periods of in the order of minutes, longer update periods would be tolerable, especially since a long battery life is also desired for the nodes.  

We can extrapolate that for a node transmission period of \SI{2.5}{\second}, the relay could forward \SI{87}{\percent} of the packets, given the same up-time and fewer packets, a behaviour similar to the one observed for \cite{8292705} (see \Cref{tab:compare}). For a duty cycle of \SI{20}{\percent} to ensure close to a year of battery life, the estimated efficiency would increase by 10 percentage points. 

The efficiency and power consumption are still subject to additional parameters such as multiple relays, tweaks to the batching policy, different values for scan and sleep times which resulting the same duty cycle, node transmission period and number of nodes. This exploration is out of the scope of this paper and left as future work.

\section{Conclusion}
\label{sec:conclusion}

We have presented an evaluation of a Bluetooth device and packet forwarding policies in mesh networks. The objective of the relay device is to extend the range of transmission between end devices, such as Bluetooth nodes, and the gateway devices, which are wall-powered and communicate with a central server. The relays allow for more area coverage without additional gateways, which are more costly, and without the necessary additional wall-power infrastructure.

We first evaluated the relay's packet reception with 2 nodes, under noise generated by 17 nodes which were not part of the system, for values of the \emph{scan window} between \SI{50}{\milli\second} and \SI{1150}{\milli\second}, and found that the relay can receive up to \SI{90}{\percent} of the node transmissions, for a node transmission period of \SI{1}{\second}. 

We then evaluated the forwarding efficiency, measured as the number of packets received by the gateway versus the total number of packets sent by the nodes. For a policy where the relay immediately forwards a received packet, only \SI{16}{\percent} of packets sent by 11 nodes are received by the gateway. By employing a policy of deferred forwarding, and multiple packet repetitions per listened node, this increases to \SI{35}{\percent}.

Finally, we measured the power draw of the device using a power analyzer, and estimated the lifetime of the four AA batteries (\SI{12000}{\milli\ampere\hour}) for different duty cycles and node transmission periods. 

\bibliographystyle{IEEEtran}
\bibliography{ble.bib}
\end{document}